\documentclass[aps,pra,reprint,showpacs,amsmath,amssymb,floatfix]{revtex4-1}

\usepackage{graphicx}
\usepackage{epsfig, float}
\usepackage{natbib}

\usepackage{bm}
\usepackage{latexsym}

\begin{document}

\title{Response of a particle in a one-dimensional lattice to an applied force: Dynamics of the effective mass}
\author{Federico Duque-Gomez}
\email[]{fduque@physics.utoronto.ca}
\author{J. E. Sipe}
\affiliation{Department of Physics, University of Toronto, Toronto, Ontario, Canada M5S1A7}
\date{\today}

\begin{abstract}
We study the behaviour of the expectation value of the acceleration of a particle in a one-dimensional periodic potential when an external homogeneous force is suddenly applied. The theory is formulated in terms of modified Bloch states that include the interband mixing induced by the force. This approach allows us to understand the behaviour of the wavepacket, which responds with a mass that is initially the bare mass, and subsequently oscillates around the value predicted by the effective mass. If Zener tunneling can be neglected, the expression obtained for the acceleration of the particle is valid over timescales of the order of a Bloch oscillation, which are of interest for experiments with cold atoms in optical lattices. We discuss how these oscillations can be tuned in an optical lattice for experimental detection.
\end{abstract}


\maketitle


\section{Introduction}

In the absence of scattering due to impurities and phonons, the wavepacket associated with a crystal electron accelerates in response to an external force $F$ (homogeneous in space and constant in time) as a particle with an effective mass. This observation is justified by the \textit{effective mass theorem} \cite{Ashcroft76}, which in the simple case of a one-dimensional lattice is 
\begin{equation}
	\left< a \right> = \frac{F}{\mathfrak{m}_{n}^{\ast}(k)}, \label{E:AccelerationUsualEffectiveMass}
\end{equation}
where $\left< a \right>$ is the expectation value of the acceleration of the wavepacket and $\mathfrak{m}_{n}^{\ast}(k)$ is inversely proportional to the curvature of the band energy,
\begin{equation}
	(\mathfrak{m}_{n}^{\ast}(k))^{-1} = \frac{1}{\hbar^2} \frac{d^{2}}{d k^2} \mathcal{E}_n(k). \label{E:UsualEffectiveMass}
\end{equation}
However, Pfirsch and Spenke \cite{Pfirsch54} argued that $\left< a \right>$ does not satisfy \eqref{E:AccelerationUsualEffectiveMass} at all times when the force is suddenly applied; instead, the initial response of the expectation value of the acceleration is characterized by the bare mass of the electron. There should be subsequent oscillations around the value calculated from the usual effective mass \eqref{E:UsualEffectiveMass}, which have a period inversely proportional to the energy gap. They die off after a characteristic time that roughly decreases with decreasing bare mass and lattice constant \cite{Pfirsch54}.\\

In typical solid-state systems, the femtosecond scale of this characteristic time and the scattering due to impurities and phonons make it difficult to observe these oscillations. Atoms in optical lattices constitute a much simpler system, where the timescale is expected to be longer and decoherence can be minimized. In these systems a constant force can be introduced by accelerating the lattice uniformly so that the atoms experience an inertial force in the lattice frame. Many experiments with ultracold atoms have been carried out to examine interesting phenomena caused by the band structure in one-dimensional optical lattices \cite{BenDahan96,Peik97,Choi99,HeckerDenschlag02,Browaeys05,Morsch06}. Here we discuss the possibility of detecting the oscillatory behaviour of the expectation value of the acceleration in ultracold atoms in a one-dimensional optical lattice when an inertial force is suddenly applied.\\

Assuming that Zener tunneling is not significant, we write an approximate semianalytical expression for the expectation value of the acceleration valid for times as long as one Bloch period (see expression \eqref{E:AccelerationFinal}). Our approach is based on constructing a wavepacket from modifed Bloch states that take into account the first order interband mixing due to the external force. These modified Bloch states are constructed using a diagonalization scheme to decouple the bands to some given order in the external force \cite{Nenciu91}, following ideas first introduced by Adams \cite{Adams56, Adams57, Adams59}, Kane \cite{Kane60} and Wannier \cite{Wannier60}. To first order, we write an explicit expression for the desired expectation value of the wavepacket acceleration. We find that at early times it reduces to the result found by Iafrate and Krieger in their study of the motion of crystal electrons shortly after a dc field is applied \cite{HessIafrate88,KriegerIafrate86, KriegerIafrate87,Iafrate98}. We discuss the validity of our perturbation approach, comparing with the results obtained from a full numerical calculation, and confirm that our approach is still valid for longer times. We identify the effect of changing the different physical parameters of the optical lattice, such as the bare mass of the atoms and the lattice constant, and find some sets of parameters that we believe would lead to experimentally measurable effects. \\

To the best of our knowledge, the type of oscillations described here have not been measured experimentally. Although the observation of this phenomenon in optical lattices is an interesting problem on its own, we believe it can also help to understand how such oscillations could be detected in solid-state systems where the femtosecond timescales are becoming more accessible through ultrafast pulses \cite{Zhu08}.\\

In Section~\ref{S:TheoreticalFramework} we summarize the strategy to calculate the expectation value of the acceleration, in terms of modified Bloch states, and illustrate its behaviour for a one-dimensional Mathieu potential with parameters adjusted to resemble those of the band structure of a semiconductor. The method used for the full numerical calcution is sketched in Section~\ref{S:FullNumericalCalculation}. In Section~\ref{S:ColdAtomsInOpticalLattices}, we show an application of the formalism to cold atoms in an optical lattice with several examples, illustrating the relevant parameters that control the oscillations. We present some conclusions in Section~\ref{S:Conclusion}.


\section{\label{S:TheoreticalFramework}THEORETICAL FRAMEWORK}

We consider a particle in an infinite one-dimensional lattice described by a wavepacket with spread in quasimomentum smaller than the extension of the Brillouin zone. In this section we describe the formalism employed to include the effect of an external homogeneous force acting on the particle.


\subsection{\label{S:HamiltonianAndModifiedBlochStates}Hamiltonian and modified Bloch states}

The Hamiltonian that describes the system is
\begin{equation}
  \mathcal{H} = \mathcal{H}_o-F(t)x, \label{E:FullHamiltonian}
\end{equation}
where $\mathcal{H}_o $ corresponds to the unperturbed part of the Hamiltonian,
\begin{equation}
  \mathcal{H}_o \equiv -\frac{\hbar^2}{2m}\frac{d^2}{dx^2} + V(x). \label{E:UnperturbedHamiltonian}
\end{equation}
The potential $V(x)$ has period $b$. We will treat $-F(t)x$ as a perturbation, but in a particular sense that will be described later.\\

If we wanted to solve this problem in the crystal momentum representation we would expand the wavefunction of the system $\Psi(x,t)$ as a wavepacket of Bloch functions, $\psi_{n}(k,x)$,
\begin{equation}
  \Psi(x,t) = \sum_n\int dk \, c_n(k,t) \psi_n(k,x), \label{E:WavepacketBlochStates}
\end{equation}
where the integral is over the Brillouin zone. The Bloch functions can be written as
\begin{equation}
  \psi_n(k,x)=\frac{1}{\sqrt{2 \pi}} u_n(k,x)e^{ikx}, \label{E:BlochFunc}
\end{equation}
where $u_n(k,x)$ has the periodicity of the one-dimensional lattice. The Bloch functions are eigenstates of the unperturbed Hamiltonian $\mathcal{H}_o$ with energies $\mathcal{E}_n(k)$. They do not diagonalize the full Hamiltonian when we include the force term, which destroys the periodicity of the system and is divergent for $|x| \rightarrow \infty$. Therefore, some care is required to express the position operator in the crystal momentum representation \cite{Blount62}
\begin{multline}
  \int_{-\infty}^{\infty}\psi_{n}^{\ast}(k,x) x \psi_{n'}(k',x) dx = \delta_{nn'}\left(-i\frac{\partial}{\partial k'} \delta(k-k') \right) \\
  + \delta(k-k')\xi_{nn'}(k),
\end{multline}
where $\xi_{nn'}(k)$ is the Lax connection \cite{Lax74}
\begin{equation}
\xi_{nn'}(k) \equiv \int_{0}^{b}u_{n}^{\ast}(k,x)i\frac{\partial}{\partial k}u_{n'}(k,x)\frac{dx}{b}.
\end{equation}
In the crystal momentum representation the Hamiltonian \eqref{E:FullHamiltonian} becomes
\begin{multline}
  \int_{-\infty}^{\infty}\psi_{n}^{\ast}(k,x) \mathcal{H} \psi_{n'}(k',x) dx = H_{nn'}(k;t) \delta(k-k') - \\
  \delta_{nn'} F(t) \cdot \left( -i \frac{\partial}{\partial k'} \delta(k-k') \right),
\end{multline}
where we have introduced the matrix elements
\begin{equation}
  H_{nn'}(k;t)\equiv \mathcal{E}_{n}^{R}(k;t) \delta_{nn'} - F(t) \cdot \xi_{nn'}(k).
\end{equation}
Here $\mathcal{E}_{n}^{R}(k;t)$ denotes the band energy renormalized by the diagonal part of the Lax connection,
\begin{equation}
  \mathcal{E}_{n}^{R}(k;t) \equiv \mathcal{E}_{n}(k) - F(t) \cdot \xi_{nn}(k). \label{E:RenormBandEner}
\end{equation}
The time evolution of the amplitudes $c_n(k,t)$ is given by the time-dependent Schr\"odinger equation
\begin{multline}
  i \hbar \frac{\partial}{\partial t} c_n(k,t) = \sum_{n'}H_{nn'}(k;t)c_{n'}(k,t) - \\
  i F(t) \cdot \frac{\partial}{\partial k}c_n(k,t). \label{E:TDSamplc}
\end{multline}\\

Now we look at the case where the force $F(t)$ is constant for all times, that is
\begin{equation}
  F(t)=F \text{, for $t \in (-\infty,\infty)$.}\label{E:ConstForce}
\end{equation}
Consider a general unitary transformation $U_{n' n}(k)$ of the Bloch states, $\psi_{n}(k,x)$,
\begin{equation}
  \phi_{n}(k,x) \equiv \sum_{n'}\psi_{n'}(k,x)U_{n' n}(k). \label{E:ModBloch}
\end{equation}
In terms of these modified Bloch states, the wavepacket \eqref{E:WavepacketBlochStates} becomes
\begin{equation}
  \Psi(x,t) = \sum_n\int dk \, b_n(k,t) \phi_n(k,x), \label{E:WavepacketModifiedBlochStates}
\end{equation}
where
\begin{equation}
  b_n(k,t) \equiv \sum_{n'} U_{n' n}^{\ast}(k) c_{n'}(k,t).
\end{equation}
From \eqref{E:TDSamplc} the evolution of $b_n(k,t)$ is given by
\begin{multline}
i \hbar \frac{\partial}{\partial t} b_n(k,t) =\sum_{n'} \bigg( \sum_{m m'} U_{mn}^{\ast}(k) H_{m m'}(k)U_{m' n'}(k) -\\
 \sum_{m}U_{mn}^{\ast}(k) i F \cdot \frac{\partial}{\partial k} U_{m n'}(k) \bigg)  b_{n'}(k,t) - i F \cdot \frac{\partial}{\partial k}b_n(k,t). \label{E:TDSamplb}
\end{multline}
In order to decouple the amplitude for the band $n$ from the rest of the bands, it would be desirable to find a transformation $U_{n' n}(k)$ that diagonalizes the term in parentheses in \eqref{E:TDSamplb},
\begin{multline}
  \sum_{m m'} U_{mn}^{\ast}(k) H_{m m'}(k)U_{m' n'}(k) -\\
 \sum_{m}U_{mn}^{\ast}(k) i F \cdot \frac{\partial}{\partial k} U_{m n'}(k) = \delta_{nn'} W_n(k). \label{E:DefU}
\end{multline}
If this equation could be solved, the solution to \eqref{E:TDSamplb} would be
\begin{equation}
  b_n(k,t) = b_n(k-\frac{1}{\hbar}F \, t,0) e^{-\frac{i}{\hbar}\int_{0}^{t} W_{n}(k-F \, (t-t')/\hbar)dt'}. \label{E:SolutionTDS}
\end{equation}\\

Wannier proved that it is possible to find a transformation that satisfies \eqref{E:DefU} using an expansion of \eqref{E:ModBloch} in powers of $F$, and described a recurrence procedure to construct it \cite{Wannier60}. Such a power series expansion is only valid if Zener tunneling between bands is not significant \cite{Wannier60, Nenciu91}. This is appropriate for the physical situations we will discuss here, where the force is small enough so that the wavepacket remains essentially in one band. In the Appendix we summarize Wannier's procedure. To first order in $F$, the unitary transformation $U_{n' n}(k)$ is
\begin{equation}
  U_{n' n}(k) \approx \delta_{n' n} + \Delta_{n'n}(k), \label{E:FirstOrderUnitaryTransf}
\end{equation}
where $\Delta_{n'n}(k)$ is a $k$-dependent dimensionless parameter that compares the interband position matrix element times the force with the energy difference between bands, $\mathcal{E}_{n' n}(k) \equiv \mathcal{E}_{n'}(k) - \mathcal{E}_{n}(k)$,
\begin{equation}
\Delta_{n'n}(k)\equiv\frac{F \cdot \xi_{n'n}(k)}{\mathcal{E}_{n' n}(k) }(1-\delta_{n'n}). \label{E:DeltaParameter}
\end{equation}
The first order approximation for $W_{n}(k)$ is
\begin{equation}
  W_{n}(k) \approx \mathcal{E}_{n}^{R}(k). \label{E:FirstOrderEnergy}
\end{equation}
Hence, the solution \eqref{E:SolutionTDS} valid to first order is \cite{Wannier60}
\begin{equation}
  b_n(k,t) \approx b_n(k - \frac{1}{\hbar}F \, t, 0) e^{-i\gamma_{n}(k - \frac{1}{\hbar}F \, t,t)}, \label{E:SolutionTDSFirstOrder}
\end{equation}
where 
\begin{equation}
  \gamma_{n}(\kappa,t) \equiv \frac{1}{\hbar} \int_{0}^{t} \mathcal{E}_{n}^{R}(\kappa + \frac{1}{\hbar}F \, t') dt' \label{E:GammaPhase}
\end{equation}
with $\kappa = k - F \, t / \hbar$.\\

Adams and Argyres \cite{Adams56, Adams57} used the modified Bloch states \eqref{E:ModBloch} with the first order approximation \eqref{E:FirstOrderUnitaryTransf},
\begin{equation}
  \phi_n(k,x)  \approx \psi_{n}(k,x) + \sum_{n'}\psi_{n'}(k,x) \Delta_{n'n}(k), \label{E:FirstOrderPhi}
\end{equation}
to construct states where the expectation value of the acceleration,
\begin{equation}
  \left< a(t) \right> \equiv \frac{d^2}{d t^2} \int_{-\infty}^{\infty} \Psi^{\ast}(x,t) \, x \, \Psi(x,t) dx,
\end{equation}
obeys the \textit{effective mass theorem} \eqref{E:AccelerationUsualEffectiveMass}. The parameter \eqref{E:DeltaParameter}, which controls the mixing between the bands, is assumed to be small but it is necessary to dress the particle in the periodic potential to establish the effective mass.\\

The expectation value of the acceleration can be written as 
\begin{equation}
  \left< a(t) \right> = \frac{F(t)}{m} + \frac{1}{i\hbar m}  \int_{-\infty}^{\infty} \Psi^{\ast}(x,t) \left[ p,\mathcal{H}_o \right] \Psi(x,t) dx, \label{E:AccelerationGen}
\end{equation}
where $p$ denotes the momentum operator and the force $F(t)$ can have any time dependence. For a wavepacket of the form
\begin{equation}
\Psi_N'(x,t)=\int dk \, b_N(k,t) \phi_N(k,x) \label{E:WavepacketEffMass}
\end{equation}
and a constant force \eqref{E:ConstForce}, the expectation value \eqref{E:AccelerationGen} becomes 
\begin{equation}
  \left< a(t) \right> = \frac{F}{m} + \int dk \, |b_N(k,t)|^2 \frac{1}{m} \mathcal{F}_{N N}(k), \label{E:Acceleration1B}
\end{equation}
where we defined 
\begin{multline}
  \mathcal{F}_{nn'}(k) \equiv \\
\frac{i}{\hbar} \sum_{m m'} U_{mn}^{\ast}(k) \, p_{m m'}(k) \mathcal{E}_{m m'}(k) \, U_{m'n'}(k) \label{E:DefPE}
\end{multline}
in terms of the momentum matrix elements
\begin{equation}
  p_{nn'}(k) \equiv \frac{2 \pi}{b} \int_{0}^{b} \psi_{n}^{\ast}(k,x) \frac{\hbar}{i} \frac{d}{dx} \psi_{n'}(k,x)dx. \label{E:MomMatElement}  
\end{equation}
To first order in $\Delta_{n'n}(k)$, $\mathcal{F}_{nn'}(k)$ is
\begin{multline}
  \mathcal{F}_{nn'}(k) \approx \frac{i}{\hbar} \bigg( \mathcal{E}_{nn'}(k)p_{nn'}(k) +\\
   \sum_{m} p_{nm}(k) \mathcal{E}_{nm}(k) \Delta_{m n'}(k) - \\
  \sum_{m} \Delta_{nm}(k)p_{mn'}(k) \mathcal{E}_{mn'}(k) \bigg). \label{E:TransfPE}
\end{multline}\\

For $n=n'=N$, equation \eqref{E:TransfPE} simplifies to
\begin{equation}
  \mathcal{F}_{N N}(k) \approx F \cdot \left( \frac{m}{\mathfrak{m}_{N}^{\ast}(k)}-1 \right), \label{E:TransfPEdiag}
\end{equation}
as a result of the well-known sum rule \cite{Lax74}
\begin{equation}
  \frac{m}{\mathfrak{m}_{n}^{\ast}(k)} = 1 + \frac{2}{m} \sum_{n' \ne n} \frac{p_{n n'}(k) p_{n'n}(k)}{\mathcal{E}_{nn'}(k)}.\label{E:EffMassSumRule}
\end{equation}
Therefore, the expectation value of the acceleration for the wavepacket \eqref{E:WavepacketEffMass} satisfies 
\begin{equation}
  \left< a(t) \right> = \int dk |b_N(k,t)|^2 \frac{F(t)}{\mathfrak{m}_{N}^{\ast}(k)}, \label{E:EffectiveMassWVP}
\end{equation}
with $\phi_N(k,x)$ in the approximation \eqref{E:FirstOrderPhi}. This corresponds to a particle accelerating with the usual effective mass \eqref{E:UsualEffectiveMass} \textit{at all times} \cite{Adams56,Adams57}.


\subsection{\label{S:ExpectationValueAccelEffMass}Expectation value of the acceleration and effective mass}

We are interested in the situation where the wavepacket is prepared initially in one band only (denoted by $N$), 
\begin{equation}
  \Psi(x,0) = \int f_N(k) \psi_{N}(k,x) dk, \label{E:InitialCondition}
\end{equation}
and the external (constant) force is suddenly applied at $t=0$, that is
\begin{equation}
  F(t) = \Theta(t) F, \label{E:ForceSudden}
\end{equation} 
where $\Theta(t)$ is the Heaviside function. The function $f_N(k)$ is assumed to have a spread smaller than the size of the Brillouin zone; for instance, in the numerical calculations presented in the next sections we will assume a Gaussian shape (centered at $k=0$)
\begin{equation}
  f_N(k) \equiv \sqrt{\frac{1}{\sigma \, \sqrt{2 \pi}}} e^{-\frac{k^2}{4 \sigma^2}}, \label{E:GaussianInitDist}
\end{equation}
characterized by the width $\sigma$.\\

For the initial wavepacket \eqref{E:InitialCondition} the term involving the commutator in \eqref{E:AccelerationGen} vanishes because the momentum matrix elements in the crystal momentum representation are diagonal in $k$,
\begin{equation}
  \int_{-\infty}^{\infty} \psi_{n}^{\ast}(k,x) \, p \, \psi_{n'}(k',x) dx= p_{nn'}(k) \delta(k-k').
\end{equation}
Hence, the particle accelerates \textit{initially} with its bare mass, $m$ \cite{Pfirsch54}. Naively, we might think that the wavepacket for $t \ge 0$ would be of the form
\begin{equation}
  \Psi_N(x,t)= \int dk \, c_N(k,t) \psi_N(k,x), \label{E:WavepacketOneBand}
\end{equation}
with $c_N(k,0)=f_N(k)$ to satisfy the initial condition \eqref{E:InitialCondition}, but then the wavepacket \eqref{E:WavepacketOneBand} would describe a particle accelerating \textit{at all times} with the bare mass. This seems to contradict the \textit{effective mass theorem} \eqref{E:AccelerationUsualEffectiveMass}; however, the contradiction is only apparent because the Bloch states of one band only are not the appropriate states to describe the particle in the presence of a homogeneous force.\\

The modified Bloch states \eqref{E:FirstOrderPhi} include the effect of the interband mixing due to the external force (to first order) and allow us to construct a wavepacket that describes a particle accelerating according to the \textit{effective mass theorem} for a constant force as shown in \eqref{E:EffectiveMassWVP}. This property suggest that such states form an appropriate basis even when the force is turned on instantaneously at $t=0$; thus, instead of a wavepacket of the form \eqref{E:WavepacketEffMass}, we use an expansion of the form \eqref{E:WavepacketModifiedBlochStates}, including the amplitudes $b_n(k,t)$ for $n \ne N$, and impose the initial condition \eqref{E:InitialCondition}, which forces the particle to respond initially with its bare mass. For later times we use the result  \eqref{E:SolutionTDSFirstOrder} to describe the evolution of the amplitudes $b_n(k,t)$ in time.\\

We find that the initial amplitudes  $b_{n}(k,t=0)$, correct to first order in $\Delta_{n'n}(k)$, are
\begin{equation}
  b_{N}(k,0) \approx f_N(k) 
 \end{equation}
and
\begin{equation}
  b_n(k,0) \approx -f_N(k) \Delta_{n N}(k), \text{ for $n \ne N$}. 
\end{equation} 
For later times we have
\begin{equation}
  b_N(k,t) \approx f_N(\kappa) e^{-i\gamma_{N}(\kappa,t)} \label{E:AmplitudeNFinal}
\end{equation}
and
\begin{equation}
  b_n(k,t) \approx -f_N(\kappa) \Delta_{n N}(\kappa) e^{-i\gamma_{n}(\kappa,t)}, \text{ for $n \ne N$}, \label{E:AmplitudeNoNFinal}
\end{equation}
where $\kappa = k - F \, t / \hbar$. Within this approximation, the amplitude for the modified Bloch state $N$ is of zeroth order while the amplitudes for the remaining modified Bloch states are of first order. The same is true for the amplitudes (correct to first order) for the usual Bloch states in the expansion \eqref{E:WavepacketBlochStates}, 
\begin{equation}
  c_{N}(k,t) \approx f_N(\kappa) e^{-i\gamma_{N}(\kappa,t)}
\end{equation}
and
\begin{multline}
  c_{n}(k,t) \approx f_N(\kappa)  \bigg( \Delta_{n N}(k) e^{-i\gamma_{N}(\kappa,t)} -  \\
 \Delta_{n N}(\kappa) e^{-i\gamma_{n}(\kappa,t)}\bigg) , \text{ for $n \ne N$}. \label{BlochAmplNoNFirstOrder}
\end{multline}
These solutions do not include Zener tunneling from the initial band $N$ to the neighbouring bands because the amplitudes in the latter remain at higher order in the expansion \cite{Wannier60, Nenciu91}.\\

Compared to \eqref{E:Acceleration1B}, there are some additional terms in the new expression for the expectation value of the acceleration \eqref{E:AccelerationGen}. For $t\ge 0$, the acceleration is
\begin{multline}
  \left< a(t) \right> = \frac{F}{m} + \sum_{n}\int dk \, |b_n(k,t)|^2 \frac{1}{m} \mathcal{F}_{n n}(k) + \\
  \sum_{\substack{n,n'\\n \ne n'}} \int dk \frac{1}{m} \mathfrak{Re} \left[ b_{n}^{\ast}(k,t) b_{n'}(k,t) \mathcal{F}_{n n'}(k) \right], \label{E:AccelerationAllOrders}
\end{multline}
where $\mathfrak{Re}[\cdot]$ denotes the real part. In the first order approximation we find that \eqref{E:AccelerationAllOrders} reduces to
\begin{multline}
  \left< a(t) \right> \approx \frac{F}{m}\int dk |f_N(\kappa)|^2 \bigg( \frac{m}{\mathfrak{m}_{N}^{\ast}(k)} +\\
    \frac{2}{m} \sum_{n \ne N} \frac{\mathcal{E}_{nN}(k)}{(\mathcal{E}_{nN}(\kappa))^2}\mathfrak{Re} \left[ p_{Nn}(k)p_{nN}(\kappa) e^{i\gamma_{Nn}(\kappa,t)}\right] \bigg), \label{E:AccelerationFinal}
\end{multline}
where we have used \eqref{E:TransfPE}, \eqref{E:TransfPEdiag}, \eqref{E:AmplitudeNFinal} and \eqref{E:AmplitudeNoNFinal}. As before we use $\kappa = k - F \, t/ \hbar$ and additionally we introduce $\gamma_{Nn}(\kappa,t) \equiv \gamma_{N}(\kappa,t) - \gamma_{n}(\kappa,t)$. The first term in \eqref{E:AccelerationFinal} describes the acceleration with the usual effective mass while the second term contains the oscillations predicted by Pfirsch and Spenke \cite{Pfirsch54}. Using the sum-rule \eqref{E:EffMassSumRule} it can verified that $\left< a(t=0) \right> = F/m$ as expected \cite{KriegerIafrate87,HessIafrate88,Iafrate98}.

\begin{figure}
\centering
\includegraphics[width=8.5cm]{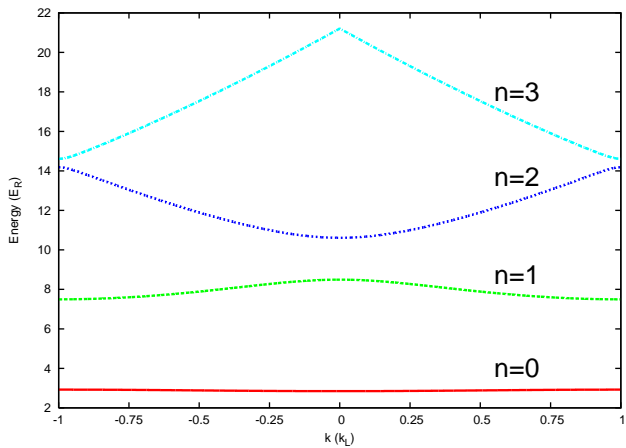}
\caption{First four energy bands ($n=0,1,2,3$) for a potential strength $s=10$. (Color online.)}
\label{fig1}
\end{figure}


\subsection{\label{S:Example}Example}

Having presented the general formalism for a one-dimensional periodic potential, we will now assume a Mathieu potential to illustrate the behaviour of the acceleration \eqref{E:AccelerationFinal}. Therefore, we set
\begin{equation}
  V(x) = V_0 \sin^2(k_Lx), \label{E:OptLattPot}
\end{equation}
where $V_0$ is a constant that determines the strength of the potential, and $k_L\equiv\pi / b$ identifies half of the Brillouin zone. The eigenvalue problem for the unperturbed Hamiltonian $\mathcal{H}_o$ can be rewritten as Mathieu's differential equation. The two linearly independent solutions of Mathieu's problem can be combined appropriately to construct Bloch-type solutions \eqref{E:BlochFunc} analytical and periodic in $k$ over the Brillouin zone $[-k_L,k_L]$ for each band index \cite{Abramowitz72,Kohn59,Shirts93a,Shirts93b}.\\

The energy scale can be characterized by the kinetic energy associated with the wavevector $k_L$,
\begin{equation}
  E_R \equiv \frac{\hbar^2 k_L^{2}}{2m}, \label{E:RecoilEnergy}
\end{equation}
and we can write $V_0 = s E_R$, where $s$ is a dimensionless constant. An example of the band structure calculated for $s=10$ is shown in Figure~\ref{fig1}. For future reference, the bands will be labeled in order of increasing energy starting with $n=0$ for the lowest band.\\

Because of the inversion symmetry of the potential $V(x)$ and the procedure used to find the Bloch functions, the diagonal part of the Lax connection appearing in the renormalized band energies \eqref{E:RenormBandEner} vanishes \cite{Kohn59} and the momentum matrix elements \eqref{E:MomMatElement} are purely real. For such a potential, \eqref{E:AccelerationFinal} simplifies to
\begin{multline}
  \left< a(t) \right> \approx \frac{F}{m}\int dk |f_N(\kappa)|^2 \bigg( \frac{m}{\mathfrak{m}_{N}^{\ast}(k)} +\\
    \frac{2}{m} \sum_{n \ne N} \frac{\mathcal{E}_{nN}(k)}{(\mathcal{E}_{nN}(\kappa))^2} p_{Nn}(k)p_{nN}(\kappa) \cos\gamma_{Nn}(\kappa,t) \bigg). \label{E:AccelerationFinalNum}
\end{multline}\\

\begin{figure}
\centering
\includegraphics[width=8.5cm]{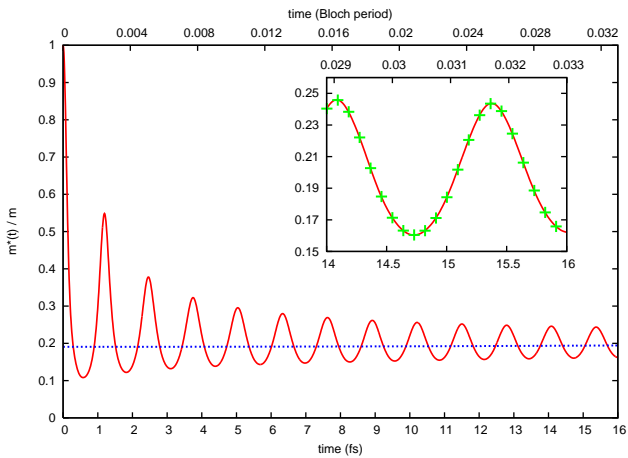}
\caption{Time dependent effective mass \eqref{E:TimeDepEffMass} for an electron wavepacket ($\sigma \approx 0.2 k_L$) initially at the center of the Brillouin zone in the band $N=2$ of a one-dimensional potential \eqref{E:OptLattPot} with $s=10$ and $b=0.5 \text{ nm}$, calculated from the first order approximation (red solid line); the blue dotted line corresponds to the usual effective mass associated with the wavepacket. For reference, the time is given both in femtoseconds and in units of the Bloch period (here $\tau_{B} \approx 485 \text{ fs}$). The first order approximation is essentially indistinguishable from the full numerical calculation result (see Section~\ref{S:FullNumericalCalculation}). We show both results for a smaller time interval in the inset, where the red solid line corresponds to the  first order approximation and the green crosses correspond to the full numerical result. The units in the inset are the same as in the main plot. (Color online.)}
\label{fig2}
\end{figure}

For times short enough after the force is applied, the variations of the momentum matrix elements and the energy differences as $\kappa$ changes with time can be neglected and we recover the expression found by Iafrate and Krieger for an electron in a dc electric field, using a vector potential gauge and accelerated Bloch states \cite{HessIafrate88,KriegerIafrate87,Iafrate98}. However, as will be seen in some examples in Section~\ref{S:ColdAtomsInOpticalLattices}, the expression \eqref{E:AccelerationFinalNum} gives a good approximation to the behaviour of the acceleration for longer times, over which the variations of the momentum matrix elements and the energy differences cannot be neglected, as long as the wavepacket remains mainly in the original band. Such long times might not be of practical interest for typical solid-state systems where it is difficult to maintain coherence over times comparable to the Bloch period, 
\begin{equation}
	\tau_{B}=h/bF,
\end{equation}
which is the time necessary for a wavepacket to return to its original position in the Brillouin zone under the action of a constant force \cite{Ashcroft76}. Experiments in optical lattices, on the other hand, can access the dynamics on timescales of the order of a Bloch period relatively easily \cite{BenDahan96,Peik97}, motivating the study of the oscillations of the effective mass as the atoms perform a full Bloch oscillation. This will be discussed in Section~\ref{S:ColdAtomsInOpticalLattices}. \\

We now illustrate some of the features contained in the result \eqref{E:AccelerationFinalNum} for an electron in a one-dimensional potential \eqref{E:OptLattPot} with parameters adjusted to resemble those of the band structure of a semiconductor; specifically, we choose $s=10$ and $b=0.5 \text{ nm}$. For this example the initial wavepacket is centered at $k=0$ in the band $N=2$, which resembles the first conduction band of a semiconductor (see Figure~\ref{fig1}). We use a  force that corresponds to a strong dc field of $1.7\times10^{7} \text{ V/m}$ and the band gap $\mathcal{E}_{21}(k=0)$ is of the order of an electron volt (here $E_R=1.5 \text{ eV}$). Using the result \eqref{E:AccelerationFinalNum} with these parameters, we plot the effective mass of the wavepacket, defined as 
\begin{equation}
  m^{\ast}(t) \equiv \frac{F}{\left< a(t) \right>}, \label{E:TimeDepEffMass}
\end{equation}
in Figure~\ref{fig2}; note that this is to be distinguished from the effective mass for an energy band \eqref{E:UsualEffectiveMass}. In this case we expect the first order approximation to be valid close to the center of the Brillouin zone.\\

This example displays the features discussed by Pfirsch and Spenke \cite{Pfirsch54}. Notice that the system behaves initially with the bare mass and, afterwards, the effective mass \eqref{E:TimeDepEffMass} oscillates around the usual effective mass \eqref{E:UsualEffectiveMass} averaged over the extension of the wavepacket.  Although there is a superposition of various oscillations in the second term of \eqref{E:AccelerationFinalNum},
\begin{multline}
	a_{\text{osc}}(t) \equiv \frac{2 F}{m^2} \sum_{n \ne N} \int dk |f_N(\kappa)|^2 \times\\
	 \frac{\mathcal{E}_{nN}(k)}{(\mathcal{E}_{nN}(\kappa))^2} p_{Nn}(k)p_{nN}(\kappa) \cos\gamma_{Nn}(\kappa,t), \label{E:aoscDefinition}
\end{multline}
associated with the different neighbouring bands, Figure~\ref{fig2} shows a distinctive oscillation with period 
\begin{equation}
	\tau_{\text{osc}} (k) = \frac{h}{|\mathcal{E}_{N\bar{n}}(k)|}, \label{E:PeriodOscillations}
\end{equation} 
given by the energy difference between the initial band, $N=2$, and its next neighbour band, $\bar{n}=1$, in the region near the center of the Brillouin zone. For the example in Figure~\ref{fig2} the period is $\tau_{\text{osc}}(0)\approx1.30 \text{ fs}$. The remaining oscillating terms have small contributions because of the much larger energy difference with respect to the band $N=2$.\\

\begin{figure}
\centering
\includegraphics[width=8.5cm]{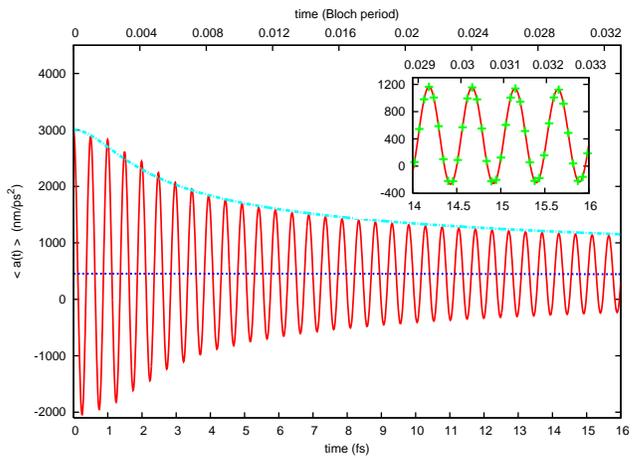}
\caption{Expectation value of the acceleration for the same electron wavepacket described in Figure~\ref{fig2} but starting in the lowest band, $N=0$. The first order approximation \eqref{E:AccelerationFinalNum} corresponds to the red solid line and the value predicted by the usual effective mass corresponds to the blue dotted line. The light blue slash-dotted line corresponds to the envelope function in \eqref{E:aoscDefinitionApproxAn}, adjusted to give the right initial value of the acceleration. As for the example shown in Figure~\ref{fig2}, there is excellent agreement between the first order approximation and  the full numerical calculation result (see Section~\ref{S:FullNumericalCalculation}). We show both results for a smaller time interval in the inset, where the red solid line correspond to the first order approximation and the green crosses correspond to the full numerical result. The units in the inset are the same as in the main plot. (Color online.)} 
\label{fig3}
\end{figure}

For the same system but with the electron starting in the lowest band, $N=0$, the acceleration is also described very well by the first order approximation \eqref{E:AccelerationFinalNum} as shown in Figure~\ref{fig3}. Compared to the previous example, the effective mass of the initial band $N=0$ in this case is larger because this band is flatter than the band $N=2$ discussed before. Due to the larger effective mass, the average value of the acceleration is smaller and the expectation value of the acceleration goes to zero several times. Therefore, it is more convenient to plot the acceleration instead of the effective mass  \eqref{E:TimeDepEffMass}. \\

The oscillations decay as a consequence of the spread of the wavepacket in $k$ \cite{Pfirsch54}. It is possible to derive a simple approximate expression for the envelope of the decaying oscillations of the acceleration while the wavepacket is moving close to the center of the Brillouin zone. If we consider only the contribution from the next neighbour band, $\bar{n}$,  and neglect the change of the momentum matrix elements and energy differences in equation \eqref{E:aoscDefinition} \cite{Pfirsch54} we can write the approximate expression
 \begin{equation}
	a_{\text{osc}}(t) \approx \frac{2 F (p_{\bar{n}N}(0))^2}{m^2 \mathcal{E}_{\bar{n}N}(0)}\int dk |f_N(k)|^2 \cos \left(\frac{1}{\hbar}\mathcal{E}_{N\bar{n}}(k)t \right). \label{E:aoscDefinitionApprox}
\end{equation}
The energy difference associated with the frequency of the oscillations can be estimated approximating the band energies close to $k=0$ by parabolas, characterized by the usual effective mass,
\begin{eqnarray}
	\mathcal{E}_{N \bar{n}}(k) &\approx& \frac{\hbar^2 k^2}{2}\left( \frac{1}{\mathfrak{m}^{*}_{N}(0)} - \frac{1}{\mathfrak{m}^{*}_{\bar{n}}(0)} \right)+  \mathcal{E}_{N \bar{n}}(0) \nonumber \\
	&=& \frac{\hbar^2 k^2}{2 \mathfrak{m}^{\text{red}}_{N \bar{n}}(0)} +  \mathcal{E}_{N \bar{n}}(0),
\end{eqnarray}
where we introduced the reduced effective mass $\mathfrak{m}^{\text{red}}_{N \bar{n}}(k)$ associated with the original band $N$ and its closest neighbour $\bar{n}$. Using this approximation and the Gaussian quasimomentum distribution \eqref{E:GaussianInitDist}, equation \eqref{E:aoscDefinitionApprox} can be evaluated analytically, yielding an oscillating term with an envelope function that controls the decay. The approximation for \eqref{E:aoscDefinition} reduces to
\begin{equation}
	a_{\text{osc}}(t) \approx \frac{2 F(p_{\bar{n}N}(0))^2\cos \left[ \frac{\mathcal{E}_{N\bar{n}}(0) t}{\hbar}  + \frac{1}{2} \tan^{-1} \left( \frac{\hbar \sigma^2 t}{\mathfrak{m}^{\text{red}}_{N \bar{n}}(0)}  \right)\right]}{m^2 \mathcal{E}_{\bar{n}N}(0) \left[ 1+\left(\frac{\hbar \sigma^2 }{\mathfrak{m}_{N \bar{n}}^{\text{red}}(0)} t \right)^2 \right]^{1/4}}. \label{E:aoscDefinitionApproxAn}
\end{equation}
The oscillations are characterized by a frequency $|\mathcal{E}_{N\bar{n}}(0)|/h$, but there is an additional contribution to the phase of the cosine in \eqref{E:aoscDefinitionApproxAn}. This contribution, however, is small in the time ranges where the approximation is appropriate. \\
 
Naturally, the approximation \eqref{E:aoscDefinitionApproxAn} is limited because we have not included the contributions from the remaining bands or the motion of the wavepacket through the Brillouin zone, but it describes appropriately the form of the decay as shown, for instance, in Figure~\ref{fig3}. We use the time it takes the oscillations in \eqref{E:aoscDefinitionApproxAn} to reduce by half,
\begin{equation}
	\tau_{\text{decay}} \equiv \frac{\sqrt{15} |\mathfrak{m}_{N \bar{n}}^{\text{red}}(0)|}{\hbar \sigma^2} , \label{E:EstimateDecay}
\end{equation}
as an estimate for the decay time of the oscillations of the acceleration around the value predicted by the usual effective mass. For the examples shown in Figures~\ref{fig2} and~\ref{fig3} the estimated decay times \eqref{E:EstimateDecay} are  $1.89 \text{ fs}$ and $4.78 \text{ fs}$, respectively, which correspond to a small fraction of the Bloch period. 


\section{\label{S:FullNumericalCalculation}FULL NUMERICAL CALCULATION}

According to the discussion in Section~\ref{S:HamiltonianAndModifiedBlochStates}, the usual effective mass can be understood as the result of a ``dressing'' process of the wavepacket in one band with small amplitudes over the neighbouring bands \cite{Adams56, Adams57}; the initial response of the wavepacket, according to the bare mass (instead of the usual effective mass), produces the oscillatory behaviour shown in expression \eqref{E:AccelerationFinalNum}. However, this simple picture is valid only if the wavepacket remains mainly in one band. The approximate solution \eqref{BlochAmplNoNFirstOrder} for the amplitudes associated with the neighbouring bands predicts a small second order correction for the probability of finding the wavepacket in those bands. Hence, situations where the population of the neighbouring bands becomes important due to Zener tunneling cannot be described within this scheme. In order to verify that the effects of the population of neighbouring bands beyond our approximate solution can be neglected in our examples, we compare the results from \eqref{E:AccelerationFinalNum} with a full numerical calculation that solves the time-dependent Schr\"odinger equation for the Hamiltonian \eqref{E:FullHamiltonian} assuming that the force has the form \eqref{E:ForceSudden} and the initial condition is given by \eqref{E:InitialCondition}.\\

For a full numerical calculation we use the split-step operator method  in its original version, where the evolution due to the kinetic energy term of the Hamiltonian is done in Fourier space \cite{Feit82,Press92}. The full Hamiltonian, given by \eqref{E:FullHamiltonian} in the position representation, corresponds to the operator $H \equiv \hat{p}^2/2m + V(\hat{x}) - F \, \hat{x}  $, where $\hat{p}$ and $\hat{x}$ are the momentum and position operators, respectively. It is convenient \cite{Peik97} to transform it to a new version
\begin{equation}
  H'(t)= S^{\dagger}(t) H S(t) - i\hbar S^{\dagger}(t)\frac{d S(t)}{dt},
\end{equation}
according to the unitary operator
\begin{equation}
  S(t) = e^{i\alpha(t) \hat{p}/\hbar}e^{-i\beta(t)\hat{x}/\hbar}e^{i \gamma(t) / \hbar}
\end{equation}
where $\alpha(t) \equiv a_L t^2 /2$, $\beta(t) \equiv m a_L t$ and $\gamma(t) \equiv m a_L^2 t^3/3$. The constant $a_L$ is set by the force, according to $F=-m a_L$. The new Hamiltonian $H'(t)$ is time dependent for $t\ge 0$ and has the same periodicity as the unperturbed Hamiltonian. It can be written in the position representation as
\begin{equation}
  \mathcal{H}'(t) = -\frac{\hbar^2}{2m}\frac{d^2}{dx^2} + s E_R \sin^2(k_L(x-\frac{1}{2}a_L \, t^2)).
\end{equation}
In the case of an electron in a dc electric field, this transformation is equivalent to a gauge transformation where the new Hamiltonian is written in terms of the vector potential, as was done by Krieger and Iafrate \cite{KriegerIafrate86, KriegerIafrate87, Iafrate98}.\\

The results presented in Figures~\ref{fig2} and~\ref{fig3}, calculated from expression \eqref{E:AccelerationFinalNum}, are in almost perfect agreement with the full numerical calculation, confirming that \eqref{E:AccelerationFinalNum} is appropriate for short times after the force has been suddenly applied. In the next section we will show that \eqref{E:AccelerationFinalNum} is also an useful estimate for longer times of the order of a full Bloch oscillation, which are relevant for experiments with cold atoms in optical lattices.

\begin{figure}
\centering
\includegraphics[width=8.5cm]{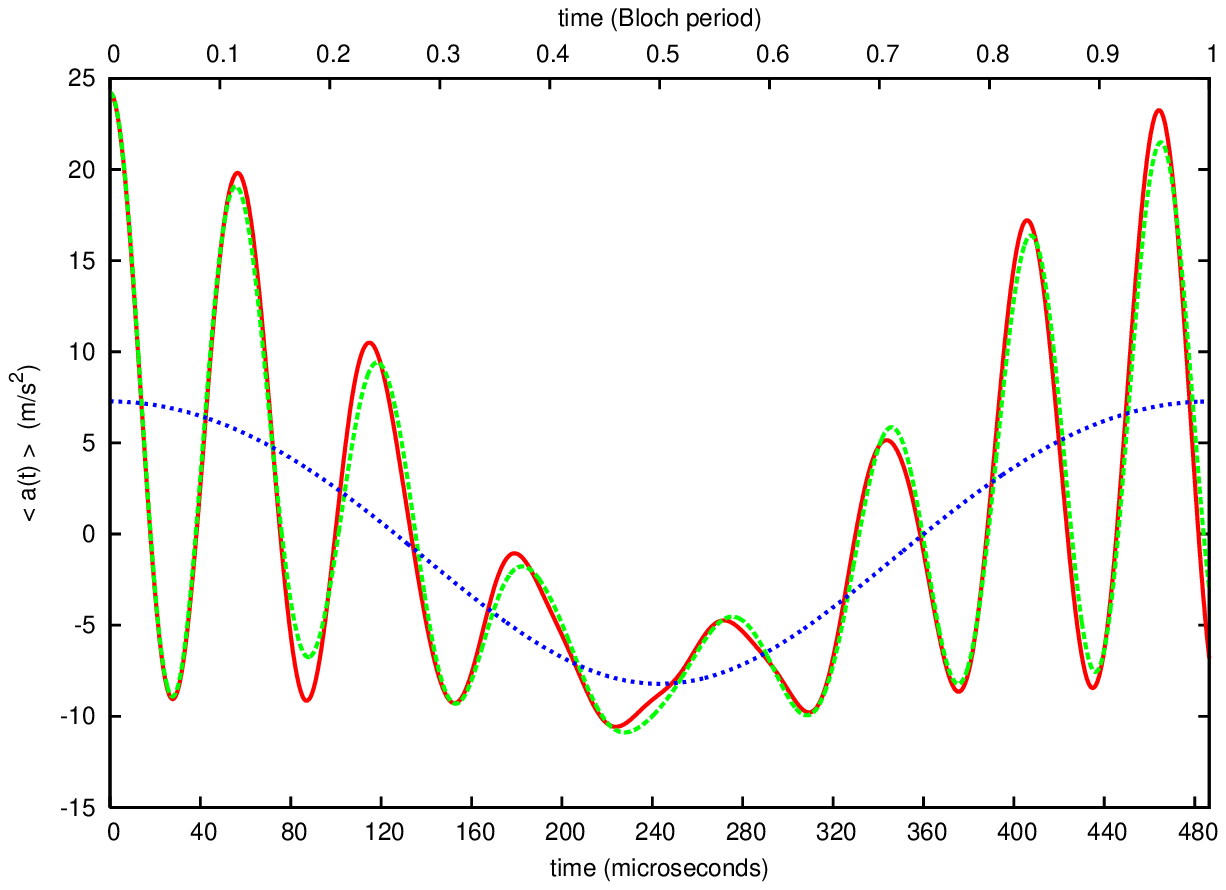}
\caption{Expectation value of the acceleration of a wavepacket ($\sigma=0.2 k_L$) initially at the center of the Brillouin zone in the band $N=0$ for a rubidium atom in an optical lattice with $s=7$ and $b=390 \text{ nm}$, calculated from the first order approximation (red solid line), the full numerical solution (green dashed line) and the usual effective mass (blue dotted line). The acceleration of the lattice is $24.2 \, m/s^2$, which corresponds to $\tilde{F}\approx0.173$ (see \eqref{E:DimlessForce}). Notice how well the approximate calculation reproduces the full numerical solution. (Color online.)}
\label{fig4}
\end{figure}


\section{\label{S:ColdAtomsInOpticalLattices}COLD ATOMS IN OPTICAL LATTICES} 

As illustrated in Section~\ref{S:Example}, the oscillations displayed in Figures~\ref{fig2} and~\ref{fig3} decay in a few femtoseconds for electron motion in a semiconductor, a timescale that is difficult to resolve experimentally. We will see that the corresponding timescales in optical lattices are of the order of microseconds. Considering the additional advantages of tunability and low decoherence that experiments in optical lattices offer, we believe that such systems are excellent candidates to study the oscillatory behaviour of the acceleration \eqref{E:AccelerationFinalNum}.\\

In experiments with cold atoms in optical lattices, a constant and homogeneous force can be introduced by uniformly accelerating the lattice with a linear increase in time of the frequency difference between the two interfering laser fields that create the lattice \cite{BenDahan96}. The appropriate Hamiltonian to describe the system in the laboratory frame is $ \mathcal{H}^{lab}(t) = \mathcal{H}'(t)$, introduced for convenience in Section~\ref{S:FullNumericalCalculation}, with $a_L$ now being the acceleration of the lattice  \cite{Peik97}.\\

\begin{figure}
\centering
\includegraphics[width=8.5cm]{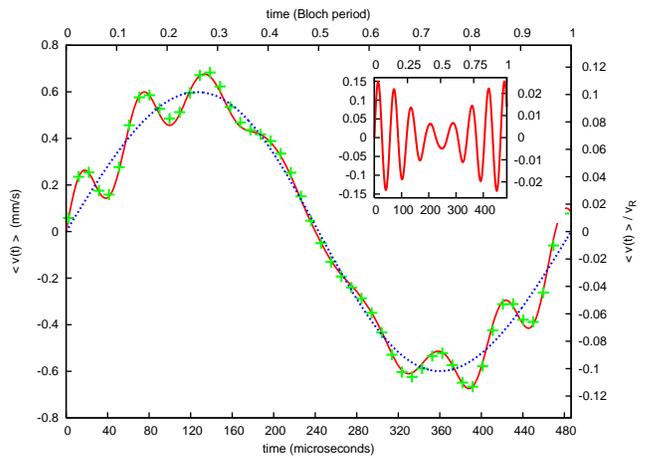}
\caption{Expectation value of the velocity of rubidium atoms in the same situation described in Figure~\ref{fig4}, calculated from the first order approximation (red solid line), the full numerical solution (green crosses) and the usual effective mass (blue dotted line). The inset shows the difference between the first order approximation and the usual effective mass prediction. The units in the inset are the same as in the main plot. (Color online.)}
\label{fig5}
\end{figure} 
 
For optical lattices the physical scales change drastically compared to the electron case described in Section~\ref{S:Example}, but the behaviour of the oscillations in \eqref{E:AccelerationFinalNum} is qualitatively the same. In order to compare the two cases, it is appropriate to consider the ratio of the initial oscillation period \eqref{E:PeriodOscillations} to the Bloch period,
\begin{equation}
	\frac{\tau_{\text{osc}}(0)}{\tau_B} = \frac{b F}{| \mathcal{E}_{N \bar{n}}(0) |} = \frac{\pi \tilde{F}}{|\tilde{\mathcal{E}}_{N \bar{n}}(0)|}, \label{E:RatioOscBloch}
\end{equation}
and the ratio of the decay time \eqref{E:EstimateDecay} to the Bloch period,
\begin{equation}
	\frac{\tau_{\text{decay}}}{\tau_B} = \frac{\sqrt{15}}{2} \frac{b F |\mathfrak{m}_{N \bar{n}}^{\text{red}}(0)|}{\pi \hbar^2 \sigma^2} = \frac{\sqrt{15}}{4} \frac{\tilde{F} |\tilde{\mathfrak{m}}_{N \bar{n}}^{\text{red}}(0)|}{\tilde{\sigma}^2}. \label{E:RatioDecayBloch}
\end{equation}
In \eqref{E:RatioOscBloch} and \eqref{E:RatioDecayBloch} we included the ratios in terms of the scaled wavevector, energy and mass:
\begin{equation}
	\tilde{k} \equiv \frac{k}{k_L} \text{, } \tilde{\mathcal{E}}_n(\tilde{k}) \equiv \frac{\mathcal{E}_n(k)}{E_R}  \text{, and } \tilde{\mathfrak{m}}^{\text{red}}_{N \bar{n}}(\tilde{k}) \equiv \frac{\mathfrak{m}^{\text{red}}_{N \bar{n}}(k)}{m}. \label{E:ScaledKEM}
\end{equation}
Note that since $\sigma$ is the spread in quasimomentum, the corresponding scaled variable is $\tilde{\sigma} \equiv \sigma / k_L$. We also introduced  the scaled force \cite{*[{A similar definition is used in }] [{. Our definition has an additional factor of $1/\pi$.} ] Zenesini09}
\begin{equation}
	 \tilde{F} \equiv \frac{b \, F}{\pi E_R}, \label{E:DimlessForce}
\end{equation}
that compares the energy drop over a unit cell, $bF$, with the characteristic energy of the system, $E_R$.\\

For the example shown in Figure~\ref{fig2} the ratios are $\tau_{\text{osc}}(0) / \tau_B \approx 0.003$ and $\tau_{\text{decay}}(0) / \tau_B \approx 0.004$, while for the example shown in Figure~\ref{fig3} they are $\tau_{\text{osc}}(0) / \tau_B \approx 0.001$ and $\tau_{\text{decay}}(0) / \tau_B \approx  0.01$.  In both electron cases the ratios are small because $\tilde{F}\approx0.002$ is also small, even though we employed a high electric field to provide the force. In contrast, for cold atoms in optical lattices it is easy to make these ratios bigger, while maintaining coherence over times of the order of a Bloch period. \\

Consider, for example, the expectation value of the acceleration calculated for rubidium atoms prepared at the center of the lowest band $N=0$ of an optical lattice with $b=390 \text{ nm}$, $s=7$ and acceleration $a_L=24.2 \, m/s^2$. The ratios \eqref{E:RatioOscBloch} and \eqref{E:RatioDecayBloch} are now larger than in the electron case  due to the increase in the force parameter to $\tilde{F}\approx0.173$. Here $\tau_{\text{osc}}(0) / \tau_B \approx 0.105$ and $\tau_{\text{decay}}(0) / \tau_B \approx 0.425$, which correspond to $\tau_{\text{osc}}(0) \approx 51.1 \text{$\mu$s}$ and $\tau_{\text{decay}}(0)\approx 207 \text{$\mu$s}$.  Figure~\ref{fig4} shows how the atom's acceleration oscillates around the behaviour expected if the atom responded with the usual effective mass at all times. Similarly to the example shown in Figure~\ref{fig3}, the acceleration goes to zero several times so we plot the acceleration instead of the time dependent effective mass \eqref{E:TimeDepEffMass}. It is important to point out that expression \eqref{E:RatioDecayBloch} does not apply here strictly speaking because in this case the decay occurs as the center of the wavepacket has moved through a considerable portion of the Brillouin zone. Nevertheless, the result \eqref{E:RatioDecayBloch} still gives a rough idea of the initial decay time (see Figure~\ref{fig4}).\\

However, one of the most striking features of Figure~\ref{fig4} is the revival of the oscillations as the wavepacket returns to $k=0$ completing a Bloch oscillation. This is not surprising when we consider expression \eqref{E:AccelerationFinalNum}, which predicts oscillations with frequencies and amplitudes periodic over the Brillouin zone. The revival is also present in the example shown in Figure~\ref{fig3} for the electron wavepacket when the acceleration is plotted for a full Bloch period. In the situation shown in Figure~\ref{fig2}, however, we cannot use equation \eqref{E:AccelerationFinalNum} to describe the behaviour of the electron wavepacket for an entire Bloch oscillation because the gap with respect to the band $n=3$ (the closest band in the region near the edge of the Brillouin zone) becomes increasingly small as the wavepacket moves closer to the band edge ($\tilde{k}=1$) and significant Zener tunneling occurs.\\  

\begin{figure}
\centering
\includegraphics[width=8.5cm]{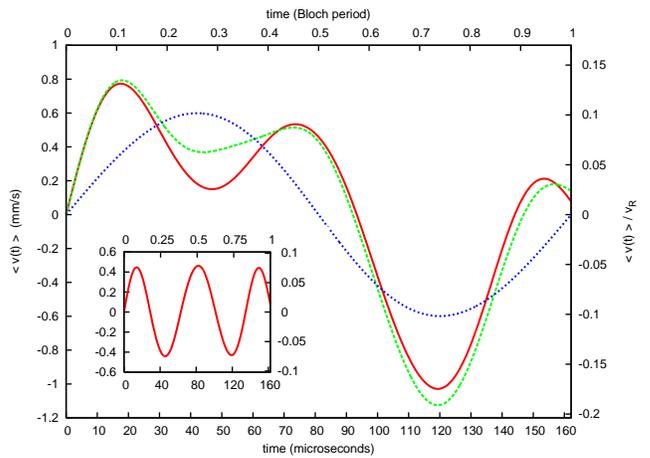}
\caption{Expectation value of the velocity for a wavepacket ($\sigma=0.2 k_L$) initially at the center of the Brillouin zone in the band $N=0$ for a rubidium atom in an optical lattice with $s=7$ and $b=390 \text{ nm}$, calculated from the first order approximation (red solid line), the full numerical solution (green dashed line) and the usual effective mass (blue dotted line). The acceleration of the lattice is $72.6  \, m/s^2$, which corresponds to $\tilde{F}\approx 0.520$. The inset shows the difference between the first order approximation and the usual effective mass prediction; this difference oscillates with a period approximately given by expression \eqref{E:RatioOscBloch}, $\tau_{\text{osc}}(0) \approx 0.315 \tau_{B} \approx 51.1 \text{$\mu$s}$. The units in the inset are the same as in the main plot. (Color online.)}
\label{fig6}
\end{figure}

In the context of cold atoms, the wavevector $k_L$ corresponds to the wavevector of the optical lattice, and the energy $E_R$, defined in \eqref{E:RecoilEnergy}, corresponds to the recoil energy, which is the energy gained (lost) by absorbing (emitting) one photon of the lattice. In \eqref{E:ScaledKEM} we used these quantities to define scaled variables for the wavevector and the energy. We can also employ them to introduce additional scaled variables for position and time: 
\begin{equation}
  \tilde{x} \equiv k_L x  \text{ and } \tilde{t} \equiv \frac{E_R}{\hbar}t. \label{E:xtkDimless}
\end{equation}
Since the acceleration has units of force over mass, the corresponding scaled acceleration is \cite{*[{A similar scaling is used in }] [{. Our definition has an additional factor of $16$.} ] Wilkinson97}
\begin{equation}
  \left< \tilde{a} (\tilde{t}) \right> = \frac{m b}{\pi E_R} \left< a (t) \right> = \frac{2 m^2 b^3}{\hbar^2 \pi^3} \left< a (t) \right>.
\end{equation}
From \eqref{E:AccelerationFinalNum} we have
\begin{multline}
  \left< \tilde{a} (\tilde{t}) \right> \approx \tilde{F}\int d\tilde{k} |\tilde{f}_N(\tilde{\kappa})|^2 \bigg( \frac{1}{\tilde{m}_{N}^{\ast}(\tilde{k})} +\\
    4 \sum_{n \ne N} \frac{\tilde{\mathcal{E}}_{nN}(\tilde{k})}{(\tilde{\mathcal{E}}_{nN}(\tilde{\kappa}))^2} \tilde{p}_{Nn}(\tilde{k}) \tilde{p}_{nN}(\tilde{\kappa}) \cos\tilde{\gamma}_{Nn}(\tilde{\kappa}, \tilde{t}) \bigg). \label{E:AccelerationDimlessFinalNum}
\end{multline}
Here we have introduced the additional scaled quantities:
\begin{eqnarray}
 \tilde{\kappa} &\equiv& \tilde{k} - \tilde{F} \, \tilde{t}, \\
 \tilde{f}_N(\tilde{\kappa}) &\equiv& \sqrt{\frac{\pi}{b}} f_N(\kappa), \label{E:GaussianDistDimless}\\
 \tilde{p}_{nn'}(\tilde{k}) &\equiv& \frac{b}{\hbar \pi} p_{nn'}(k), \\
 \tilde{\gamma}_{Nn}(\tilde{\kappa}, \tilde{t}) &\equiv& \gamma_{Nn}(\kappa,t).
\end{eqnarray}\\

In many experiments that study the motion of cold atoms in optical lattices, the measured variable is the velocity of the wavepacket rather than its acceleration \cite{BenDahan96,Peik97,Choi99,HeckerDenschlag02,Browaeys05,Morsch06}.  In Figure~\ref{fig5} we show a plot of the velocity for the same parameters as Figure~\ref{fig4}. As before, we include a plot of the prediction based on the usual effective mass, which clearly shows a full Bloch oscillation. Note also the good agreement between the prediction based on the first order calculation discussed in section \ref{S:TheoreticalFramework} and the full numerical solution, even for long times after the force was suddenly applied for both the acceleration (Figure~\ref{fig4}) and the velocity (Figure~\ref{fig5}).\\

\begin{figure}
\centering
\includegraphics[width=8.5cm]{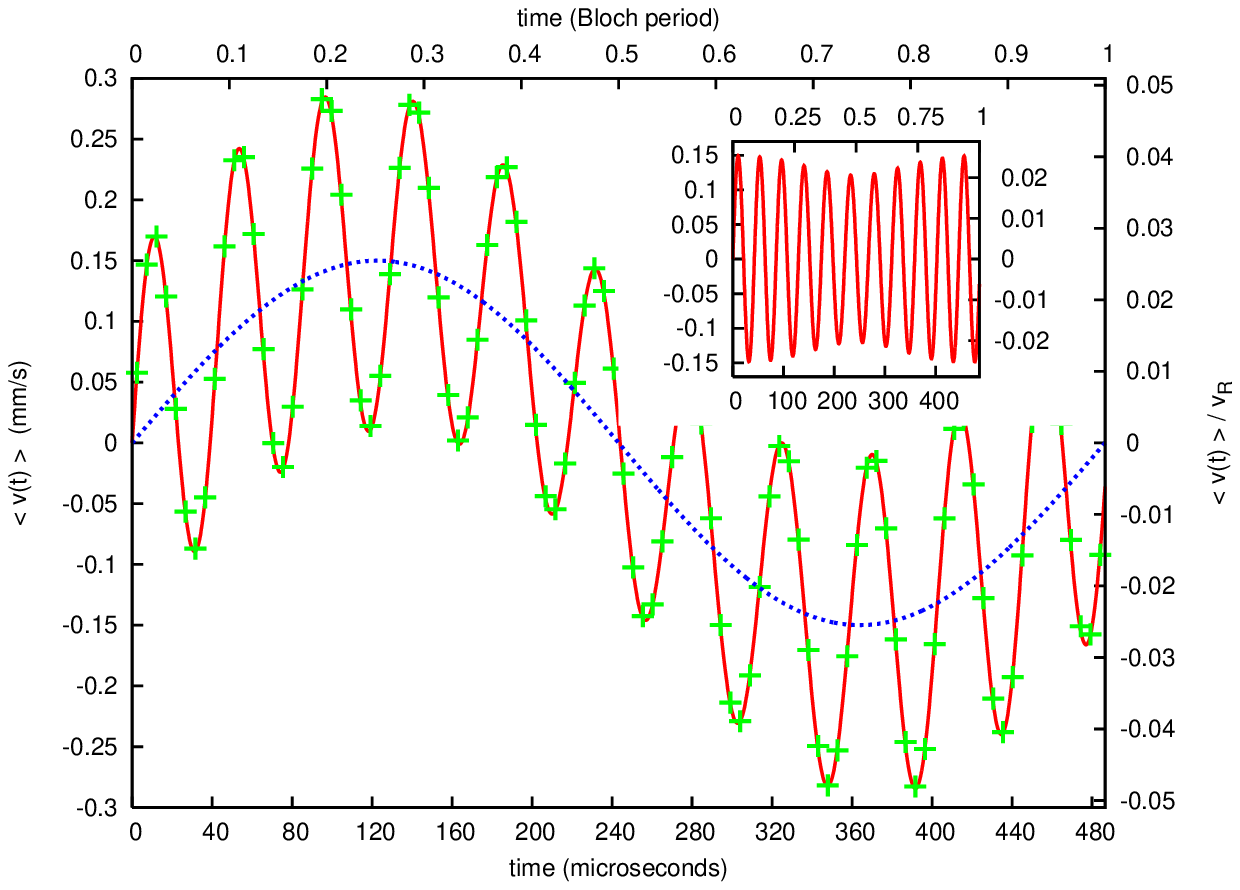}
\caption{Expectation value of the velocity for a wavepacket ($\sigma=0.2 k_L$) initially at the center of the Brillouin zone in the band $N=0$ for a rubidium atom in an optical lattice with $s=13$ and $b=390 \text{ nm}$, calculated from the first order approximation (red solid line), the full numerical solution (green crosses) and the usual effective mass (blue dotted line). The acceleration of the lattice is $24.2 \, m/s^2$, which corresponds to $\tilde{F}\approx0.173$. The inset shows the difference between the first order approximation and the usual effective mass prediction; this difference oscillates with a period approximately given by expression \eqref{E:RatioOscBloch}, $\tau_{\text{osc}}(0) \approx 0.0859 \tau_{B} \approx 41.8 \text{$\mu$s}$. The units in the inset are the same as in the main plot. (Color online.)}
\label{fig7}
\end{figure}

In the rest of this section we discuss the behaviour of the velocity in various situations realizable in experiments with cold atoms, where the natural scale for the velocity is given by the recoil velocity,
\begin{equation}
  v_R \equiv \frac{\hbar \pi}{mb}. \label{E:DefRecoilVel}
\end{equation} 
Accordingly, we use the scaled velocity 
\begin{equation}
  \left< \tilde{v}(\tilde{t}) \right>=\frac{\left<v(t)\right>}{v_R}, \label{E:DefScaledVel}
\end{equation}
which results from integrating the dimensionless acceleration,
\begin{equation}
 \left< \tilde{v}(\tilde{t}) \right> \equiv \int_{0}^{\tilde{t}}d\tilde{t}' \left< \tilde{a}(\tilde{t}') \right>.
\end{equation}
Note that we assumed the wavepacket's initial velocity is zero because, it starts at the center of the Brillouin zone in all the examples considered here.\\

The deviations of the actual expectation value of the velocity from the usual effective mass approximation value are controlled by the amplitude of the oscillations of the scaled acceleration \eqref{E:AccelerationDimlessFinalNum} and the recoil velocity \eqref{E:DefRecoilVel}. First, we fix the product of the mass and lattice constant to that used in the previous example (rubidium atoms with $b=390 \text{ nm}$) and explore the effect of changing $\left< \tilde{a} (\tilde{t}) \right>$. The recoil velocity in this case is $v_R\approx 5.89 \text{ mm/s}$.\\

 According to expression \eqref{E:AccelerationDimlessFinalNum},  $\left< \tilde{a} (\tilde{t}) \right>$ for a given initial band  $N$ depends only on the width of the quasimomentum distribution; the strength of the periodic potential, characterized by $s$; and the scaled force $\tilde{F}$. For fixed values of these three parameters the oscillations preserve their shape and are only rescaled when the physical parameters, such as the bare mass or the lattice constant, are changed.\\ 

The width of the quasimomentum distribution and the reduced effective mass control how fast the oscillations decay according to \eqref{E:RatioDecayBloch}. However, as pointed out before, the estimate \eqref{E:RatioDecayBloch} does not take into account the motion of the wavepacket through the Brillouin zone over one Bloch period since it was derived from the approximation \eqref{E:aoscDefinitionApproxAn}, which misses completely the revival of the oscillations of the effective mass. The decay can be minimized or even removed if the oscillations of the effective mass do not have time to decay before the periodicity of the terms in \eqref{E:AccelerationFinalNum} (over one Bloch period) returns the amplitude of the oscillations to the initial value.\\

\begin{figure}
\centering
\includegraphics[width=8.5cm]{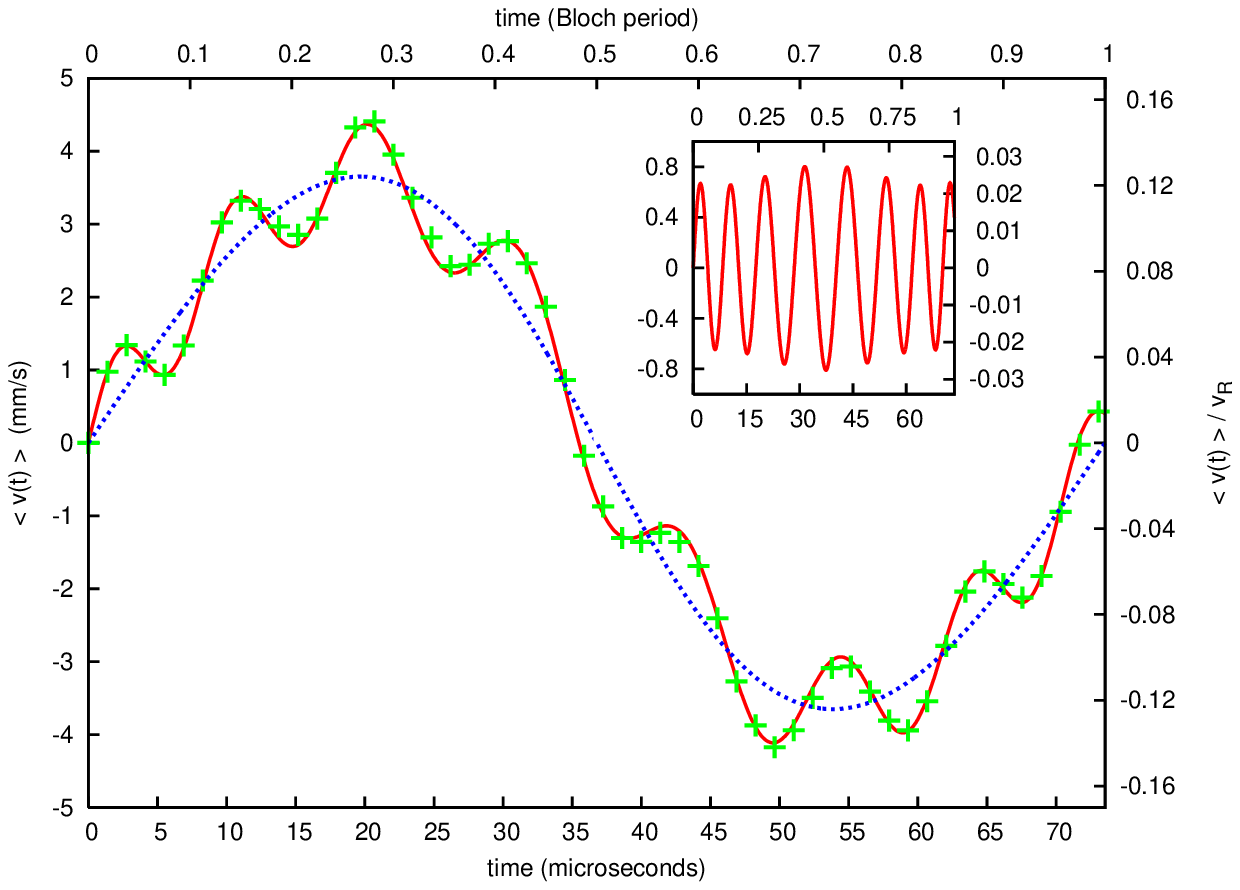}
\caption{Expectation value of the velocity for a wavepacket ($\sigma=0.004 k_L$, as has been achieved in \cite{Ryu06}) initially at the center of the Brillouin zone in the band $N=0$ for a sodium atom in an optical lattice with $s=7$ and $b=295 \text{ nm}$, calculated from the first order approximation (red solid line), the full numerical solution (green crosses) and the usual effective mass (blue dotted line). The acceleration of the lattice is $800 \, m/s^2$, which corresponds to $\tilde{F}\approx0.173$. The inset shows the difference between the first order approximation and the usual effective mass prediction; this difference oscillates with a period approximately given by expression \eqref{E:RatioOscBloch}, $\tau_{\text{osc}}(0) \approx 0.105 \tau_{B} \approx 7.73 \text{$\mu$s}$. The units in the inset are the same as in the main plot. (Color online.)}
\label{fig8}
\end{figure}

From \eqref{E:RatioDecayBloch} we would expect the decay time to become comparable to the Bloch period when increasing the scaled force $\tilde{F}$ and the reduced effective mass $\tilde{\mathfrak{m}}^{\text{red}}_{N \bar{n}}(0)$. The first case is illustrated in Figure~\ref{fig6} where we show how the increase of $\tilde{F}$, keeping the other parameters fixed, eliminates the decay shown in Figure~\ref{fig5}. The second case is shown in Figure~\ref{fig7} where, instead of changing $\tilde{F}$, we increase $s$ in the parameters used for Figures~\ref{fig4} and~\ref{fig5} so that the bands $N=0$ and $\bar{n}=1$ become flatter and the reduced effective mass increases. In both Figure~\ref{fig6} and Figure~\ref{fig7} the spread of the wavepacket is the same as for Figures~\ref{fig2}-\ref{fig5}. We can control the decay by changing the spread of the wavepacket as shown in Figure~\ref{fig8}, where we plot our results for sodium atoms with $\tilde{F} \approx 0.173$ and $s=7$ (as in Figures~\ref{fig4} and~\ref{fig5}) for a very small spread of the initial wavepacket ($\tilde{\sigma}=0.004$). Notice that in this case, instead of a decay of the oscillations after half a Bloch period, there is a slight increase of the amplitude of the oscillations due to the higher sensitivity of the wavepacket to the changes of the momentum matrix elements and the energy differences at the center of the wavepacket as it moves through the Brillouin zone. Such small width in quasimomentum can be achieved experimentally in Bose Einstein condensates of sodium atoms \cite{Ryu06}. \\

The parameters $\tilde{F}$ and $s$ also control the amplitude of the oscillations of $\left< \tilde{a} (\tilde{t}) \right>$ and $\left< \tilde{v} (\tilde{t}) \right>$. From expression \eqref{E:AccelerationDimlessFinalNum} it is clear that the amplitudes scale linearly with $\tilde{F}$  (compare Figures~\ref{fig5} and \ref{fig6}).  In the range of $s$ values explored here (from $s=7$ to $s=14$), the effect of modifying $s$ on the amplitude of the oscillations is much smaller than the effect of $\tilde{F}$ (compare Figures~\ref{fig5}, \ref{fig6} and \ref{fig7}). Thus, in an experimental setting, the oscillations can be made more visible by increasing the force within the limits where Zener tunneling is not significant.\\

\begin{figure}
\centering
\includegraphics[width=8.5cm]{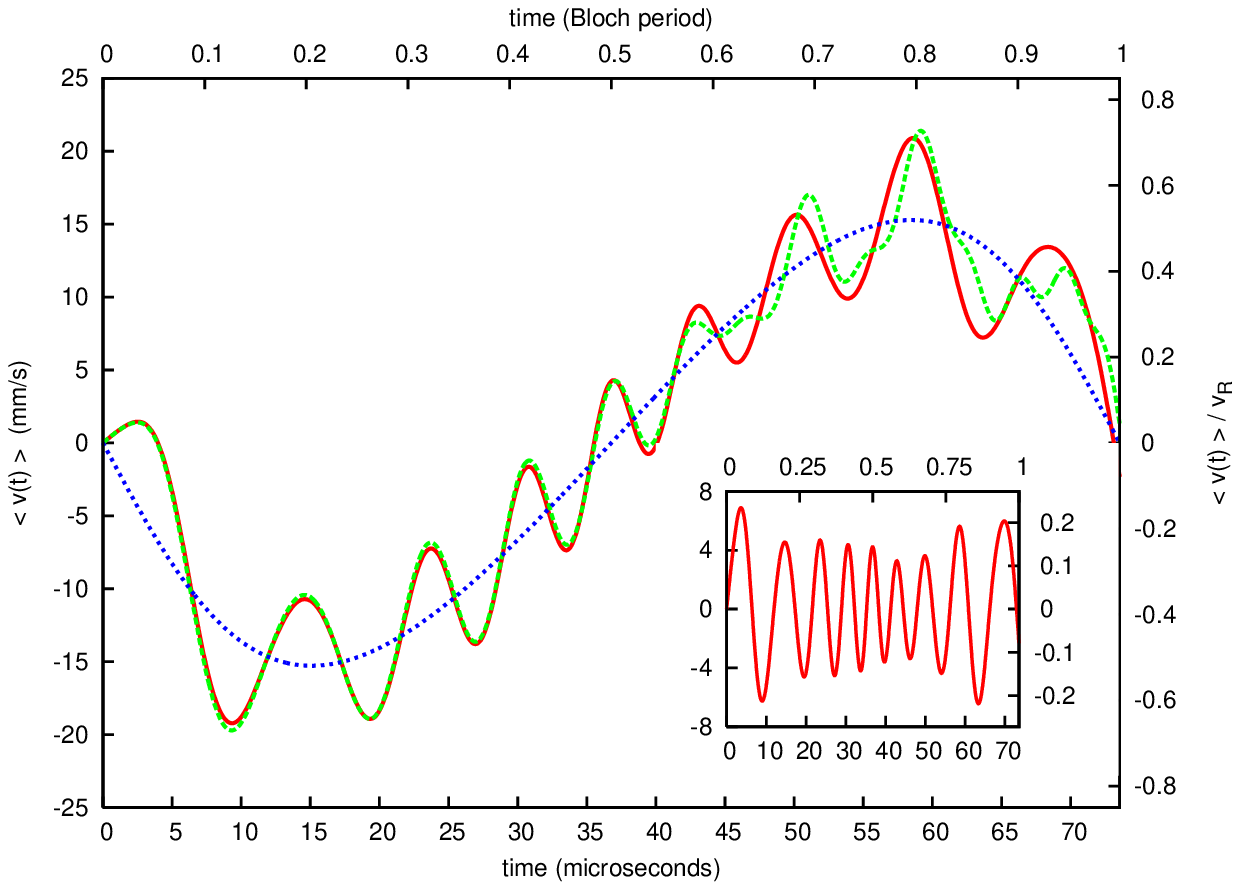}
\caption{Expectation value of the velocity of a wavepacket ($\sigma=0.01 k_L$) initially at the center of the Brillouin zone in the band $N=1$ for a sodium atom in an optical lattice with $s=13$ and $b=295 \text{ nm}$, calculated from the first order approximation (red solid line), the full numerical solution (green dashed line) and the usual effective mass (blue dotted line). The acceleration of the lattice is $800 \, m/s^2$, which corresponds to $\tilde{F}\approx0.173$. Parameters are as in \cite{Browaeys05}. The inset shows the difference between the first order approximation and the usual effective mass prediction. The oscillations shown in the inset start with a period approximately given by expression \eqref{E:RatioOscBloch}, $\tau_{\text{osc}}(0) \approx 0.176 \tau_{B} \approx 12.9 \text{$\mu$s}$; then they squeeze as the wavepacket moves through the edge $k=k_L$ of the Brillouin zone (where the gap between bands $N=1$ and $n=2$ increases); and finally they return to the starting period.  The units in the inset are the same as in the main plot. (Color online.)}
\label{fig9}
\end{figure}

\begin{figure}
\centering
\includegraphics[width=8.5cm]{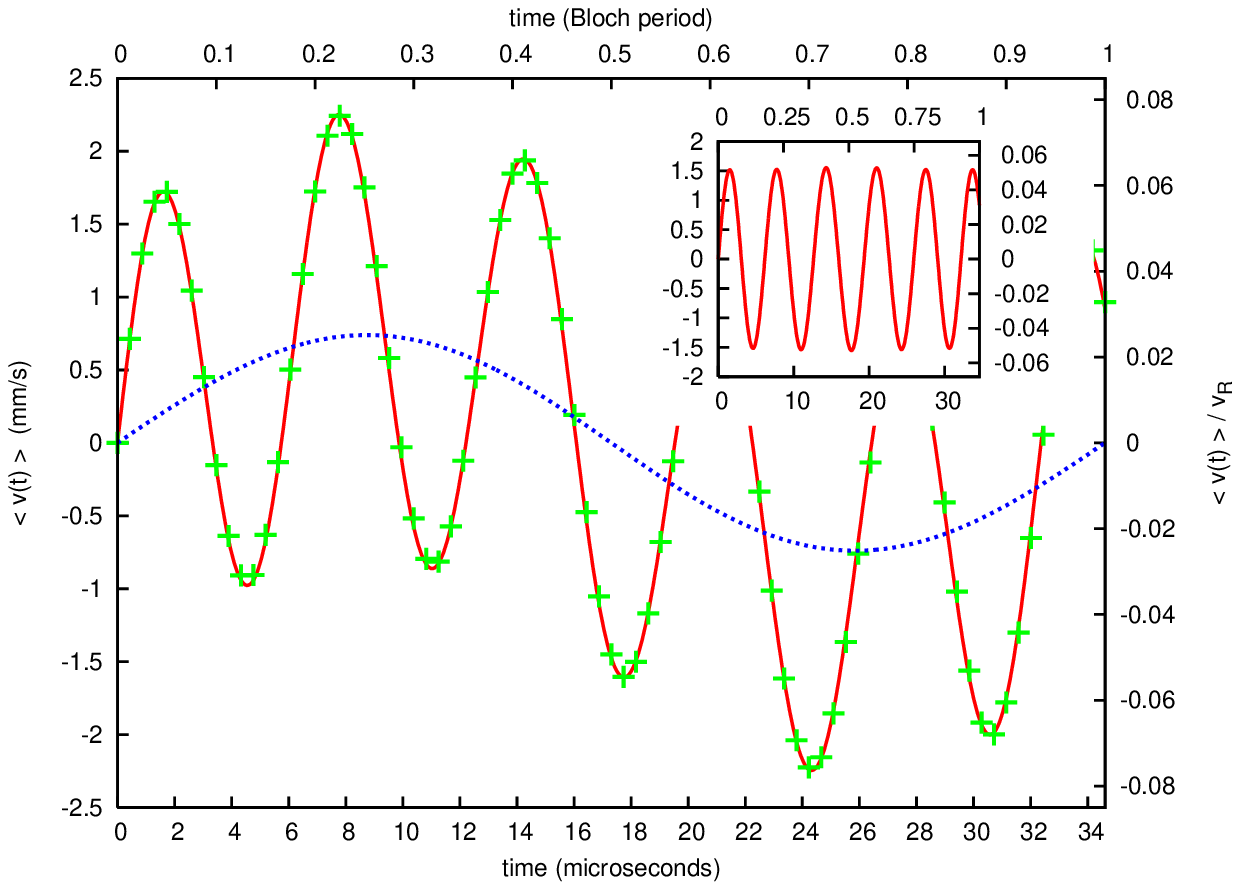}
\caption{Expectation value of the velocity for a wavepacket ($\sigma=0.01 k_L$) initially at the center of the Brillouin zone in the band $N=0$ for a sodium atom in an optical lattice with $s=14$ and $b=295 \text{ nm}$ , calculated from the first order approximation (red solid line), the full numerical solution (green crosses) and the usual effective mass (blue dotted line). The acceleration of the lattice is $1700 \, m/s^2$, which corresponds to $\tilde{F}\approx0.369$. Parameters are as in \cite{HeckerDenschlag02}. The inset shows the difference between the first order approximation and the usual effective mass prediction; this difference oscillates with a period approximately given by expression \eqref{E:RatioOscBloch}, $\tau_{\text{osc}}(0) \approx 0.176 \tau_{B} \approx 6.10 \text{$\mu$s}$. The units in the inset are the same as in the main plot. (Color online.)}
\label{fig10}
\end{figure}

The velocity deviations tend to be larger for lighter atoms and smaller lattice constants since the recoil velocity is larger. For instance, the expectation value of the velocity calculated for sodium atoms in an optical lattice with $b=295 \text{ nm}$ for the same $s$, $\tilde{\sigma}$ and $\tilde{F}$ as in Figure~\ref{fig5} is a simple rescaled version of the results in that figure according to the new recoil velocity of the system ($v_R \approx 29.4 \text{ mm/s}$). Some other examples of calculations for sodium atoms are shown in Figures~\ref{fig9} and~\ref{fig10} with parameters previously used in experiments that investigate the acceleration of Bose Einstein condensates \cite{Browaeys05, HeckerDenschlag02}. The detection of the oscillations shown here imply resolving deviations from the usual Bloch oscillation that are small compared to the recoil velocity and probably  within the uncertainty in the measurements in those references \footnote{The predicted oscillations are even smaller for other experiments with rubidium atoms such as the ones in \cite{BenDahan96} and \cite{Peik97}.}.\\

Figure~\ref{fig9} illustrates the case where the initial band is $N=1$ instead of the ground state band used in all the other examples for cold atoms.  An important difference with respect to the case with $N=0$ is the increase of the amplitude of the oscillations. Compare, for instance, the insets of Figures~\ref{fig7} and~\ref{fig9} in terms of the scaled velocity \eqref{E:DefScaledVel}. Although Figure~\ref{fig7} corresponds to rubidium atoms, in scaled units it is equivalent to a plot of the velocity for sodium atoms with the same $s=13$ and $\tilde{F}\approx0.173$ but with the wavepacket starting in the band $N=0$ with spread $\tilde{\sigma}=0.2$. One of the reasons for this increase is the smaller gap between bands $N=1$ and $n=2$ at $k=0$. Both figures predict almost no decay, but the reasons are different. For the situation in Figure~\ref{fig7} the cause is the small curvature of the band $N=0$, which increases the reduced effective mass associated with $N=0$ and $\bar{n}=1$. In Figure~\ref{fig9}, on the other hand, the cause is the small spread in quasimomentum and not the curvature of the bands; in fact, the larger magnitudes of  the curvatures of $N=1$ and $\bar{n}=2$ make the absolute value of the reduced effective mass associated with them smaller, and therefore the decay would be faster if the spread $\tilde{\sigma}$ were the same as for Figure~\ref{fig7}.\\ 

The last example, shown in Figure~\ref{fig10}, corresponds to a situation where the wavepacket is again in the lowest band, $N=0$, but under a force that is approximately two times larger than the one used in previous examples with sodium atoms and produces an acceleration $a=1700 \, m / s^2$; the strength of the potential is set to $s=14$. Accordingly, the amplitude of the oscillations of the velocity is approximately doubled compared to the cases shown in Figure~\ref{fig8} and Figure~\ref{fig7} (for the latter the comparison is in terms of the scaled velocity). The first order approximation works very well due to the large gap between the bands $N=0$ and $\bar{n}=1$ for $s=14$ and the deviations with respect to the full numerical shown in Figure~\ref{fig6} ($s=7$) when increasing the force do not occur. In Figure~\ref{fig9} using $s=13$ is not enough to prevent deviations with respect to the full numerical solution because the gap between the bands $N=1$ and $\bar{n}=2$ at $k=0$ is smaller than the energy difference between bands $N=0$ and $\bar{n}=1$.\\

The validity of the first order approximation \eqref{E:AccelerationDimlessFinalNum}, relies on the populations in the bands $n \ne N$ being small because, according to \eqref{BlochAmplNoNFirstOrder}, there is no population of these bands to first order in $\Delta_{nN}(k)$. In all the examples shown here the probabilities in the bands $n \ne N$ do not exceed 2\%, justifying the good agreement in most of the cases. The largest deviations occur for Figures~\ref{fig6} and~\ref{fig9}, although in both cases the first order approximation describes correctly the overall behaviour.\\ 

\begin{figure}
\centering
\includegraphics[width=8.5cm]{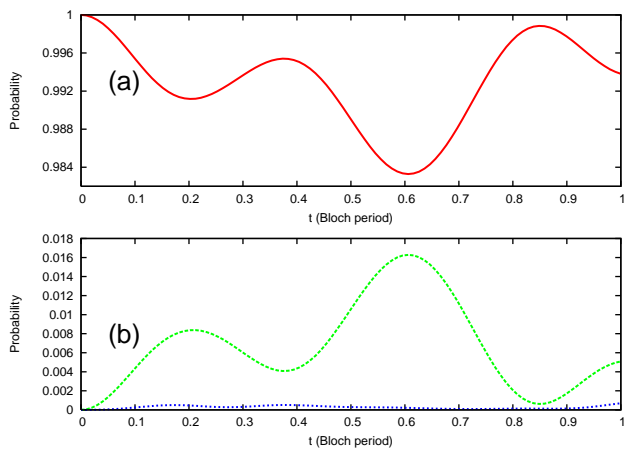}
\caption{Probability of the rubidium atoms in the optical lattice described in Figure~\ref{fig6} to be in (a) the initial band $N=0$ (red solid line) and (b) the next two neighbouring bands: $\bar{n}=1$ (green dashed line) and $n=2$ (blue dotted line). All the populations were calculated using the full numerical solution. Notice that the probability for the band $n=2$ ($\lesssim 0.07\%$) is significantly smaller than for the other two bands; we have not included the populations for higher bands since they are even smaller. The time is in units of the Bloch period. (Color online.)} 
\label{fig11}
\end{figure}

Figure~\ref{fig11} shows the populations of various bands as they change in time for the case presented in Figure~\ref{fig6}. Notice that near the middle of the Bloch oscillation the population of the band $\bar{n}=1$ overshoots the population at the end of the Bloch cycle. This overshooting has been discussed before (\cite{Holthaus00}, and references therein) when studying Zener tunneling for much stronger forces. The additional oscillations in the population observed in our case are related to the oscillations of the expectation value of the acceleration (and the velocity) around the prediction from the usual effective mass. Only a certain combination of amplitudes in the different bands yields a wavepacket behaving with the usual effective mass as discussed in Section~\ref{S:HamiltonianAndModifiedBlochStates} (see expression \eqref{E:WavepacketEffMass}); thus, we would expect that the oscillations of the population are accompanied by oscillations in the dynamical response around the usual effective mass prediction.\\

\begin{figure}
\centering
\includegraphics[width=8.5cm]{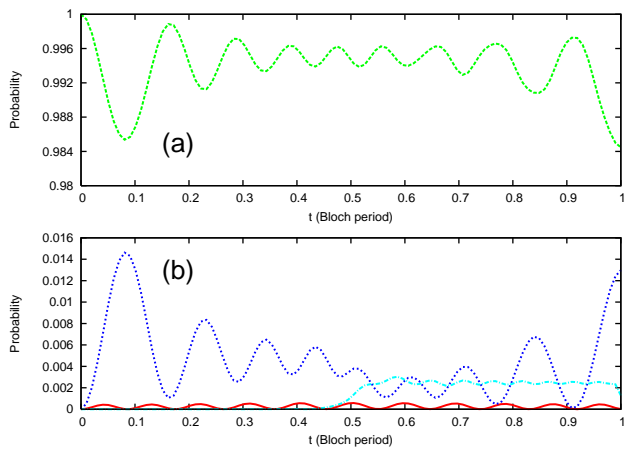}
\caption{Probability of the sodium atoms in the optical lattice described in Figure~\ref{fig9} to be in (a) the initial band $N=1$ (dashed green line) and (b) the next three neighbouring bands: $n=0$ (red solid line), $n=2$ (blue dotted line) and $n=3$ (light blue slash-dotted line). All the populations were calculated using the full numerical solution. We have not included the populations for higher bands since they are smaller than the populations for the bands shown here. The time is in units of the Bloch period. (Color online.)} 
\label{fig12}
\end{figure}

The connection between the oscillations in the population and the oscillations of the effective mass is more evident after comparing Figures~\ref{fig11} and~\ref{fig12}. The latter shows the oscillations of the populations for the case discussed in Figure~\ref{fig9}, which are faster as expected for faster oscillations of the velocity shown in Figure~\ref{fig9} compared to Figure~\ref{fig6}. This relation between the oscillations of the population and the dynamical response is found in all the other examples, where the amplitudes are smaller than those shown in Figures~\ref{fig11} and \ref{fig12}.\\

Figure~\ref{fig12} also explains the discrepancy between the full numerical solution and the first order approximation in Figure~\ref{fig9}, near the end of the Bloch period. Compared to the example with rubidium atoms in Figure~\ref{fig6}, where the main deviation occurs for the amplitude of the oscillations, the deviations in Figure~\ref{fig9} clearly show a superposition of faster oscillations. Such faster oscillations are a clear indication that the population of the neighbouring bands $n = 0 $ and $n \ge 3$ for the example in Figure~\ref{fig9} are more important than the population of the bands $n \ge 2$ for the example in Figure~\ref{fig6}. Consistent with this observation, the probability in Figure~\ref{fig12} for the band $n = 3$ ($\lesssim 0.30 \%$) is larger than the probability in Figure~\ref{fig11} for the band $n = 2$  ($\lesssim 0.07 \%$). 

\section{\label{S:Conclusion}CONCLUSION} 

In summary, we have studied the effect of an external homogeneous force acting on a particle prepared initially in one band of a one-dimensional periodic lattice when the force is suddenly applied and remains constant afterwards. As predicted by Pfirsch and Spenke \cite{Pfirsch54}, the expectation value of the acceleration responds initially with the bare mass, and oscillates around the value predicted using the usual effective mass.\\ 

Using the perturbation scheme of Adams and Wannier \cite{Adams56, Wannier60} for a wavepacket located initially in one band only, we derived an expression for the expectation value of the acceleration that is valid over a full Bloch oscillation, provided that Zener tunneling is not significant. In this picture, the wavepacket responds with the usual effective mass as it acquires certain small components of Bloch functions from neighbouring bands, but due to the initial condition, which requires the wavepacket to respond with the bare mass, the expectation value of the acceleration \eqref{E:AccelerationFinal} has various terms oscillating around the usual effective mass prediction \eqref{E:EffectiveMassWVP} with different frequencies. For the cases considered here, the most important contribution in the sum over the different bands in \eqref{E:AccelerationFinal}  comes from the closest band to the initial one; accordingly, the frequency of the oscillations is governed by the energy difference between these two bands as the wavepacket moves through the Brillouin zone. The oscillations can decay because of the spread of the wavepacket in quasimomentum, but the periodicity of the momentum matrix elements and energy differences in \eqref{E:AccelerationFinal} produces a revival of the oscillations as the wavepacket completes a  full Bloch oscillation. \\

We presented calculations for a toy model of a one-dimensional semiconductor illustrating the features described by Pfirsch and Spenke. The initial decay in this case is very fast (femtosecond timescale) and occurs while the wavepacket has moved over a small portion of the Brillouin zone, allowing us to write a simple expression for the envelope function that controls the decay of the oscillations according to the predictions by Pfirsch and Spenke (see  expression \ref{E:aoscDefinitionApproxAn}).\\ 

We also showed an analysis of the oscillations in a system of cold atoms in an optical lattice, where the timescales of the oscillations and the decay are much longer (of the order of microseconds) and comparable to a Bloch period. We analysed the effects of tuning the different parameters, such as the force, the strength of the potential, the spread of the wavepacket in quasimomentum, the bare mass of the atom and the lattice constant. Since the velocity of the atoms is a good candidate for experimental measurements, we plotted its expectation value, showing how the deviations from the usual effective mass prediction can be comparable to the oscillations calculated from the usual effective mass alone. Since the decoherence in optical lattices is much smaller than in typical solid-state systems, it would be possible, in principle, to detect the oscillations during times of the order of a Bloch oscillation. In the case of optical lattices, the decay can be easily minimized or even suppressed when the revival of the oscillations of the effective mass is faster than the decay due to the spread of the wavepacket in quasimomentum. This feature could be exploited to determine how much decoherence occurs during one Bloch oscillation.


\begin{acknowledgements}
This work was supported by the Natural Sciences and Engineering Research Council of Canada (NSERC). We thank Aephraim Steinberg, Chao Zhuang, Matin Hallaji, Alex Hayat and other members of Aephraim Steinberg's research group for valuable discussions.
\end{acknowledgements}


\appendix*

\section{Wannier's procedure}

In this appendix we sketch the method developed by Wannier to decouple the bands to any order in the force $F$ \cite{Wannier60}. Equation \eqref{E:DefU} can be rewritten as
\begin{multline}
  \left[ \mathcal{H}_o - F\cdot \left(x+ i \frac{\partial}{\partial k} \right) \right] \phi_n(k,x) = W_n(k) \phi_n(k,x),\label{E:PseudoEigenvalue}
\end{multline}
for the modified Bloch states \eqref{E:ModBloch}. Notice that equation \eqref{E:PseudoEigenvalue} takes the form of an eigenvalue problem, but with the peculiarity that both the operator (acting on $\phi_n(k,x)$) and $W_n(k)$  depend on $k$.\\

The parameter $\tilde{F}$ introduced in \eqref{E:DimlessForce} is appropriate to characterize how strong the external force is with respect to the lattice potential. For convenience, we use the other dimensionless variables defined in \eqref{E:ScaledKEM} and \eqref{E:xtkDimless}, and accordingly we introduce $\tilde{\mathcal{H}}_o \equiv \mathcal{H}_o/E_R $ and $\tilde{W}_n(\tilde{k}) \equiv W_n(k)/E_R$. The Bloch states are kept unchanged so we can write  $\tilde{\psi}_n(\tilde{k},\tilde{x}) \equiv \psi_n(k,x)$  in the new variables. The same is assumed for the unitary transformation, $\tilde{U}_{n' n}(\tilde{k}) \equiv U_{n' n}(k)$,  and therefore $\tilde{\phi}_n(\tilde{k},\tilde{x}) \equiv \phi_n(k,x)$.\\

With these definitions \eqref{E:PseudoEigenvalue} can be rewritten as
\begin{equation}
  \left[ \tilde{\mathcal{H}}_o - \tilde{F}\cdot \left(\tilde{x}+ i \frac{\partial}{\partial \tilde{k}} \right) \right] \tilde{\phi}_n(\tilde{k},\tilde{x}) = \tilde{W}_n(\tilde{k}) \tilde{\phi}_n(\tilde{k},\tilde{x}). \label{E:PseudoEigenvalueTransf}
\end{equation}
We attempt to solve this equation expressing  
\begin{equation}
  \tilde{U}_{n' n}(\tilde{k}) \approx \sum_{\nu} \tilde{\mathcal{U}}_{n' n}^{(\nu)}(\tilde{k}) \, \tilde{F}^{\nu} \label{E:TransfExp}
 \end{equation}
and
\begin {equation} 
  \tilde{W}_n(\tilde{k}) \approx \sum_{\nu} \tilde{\mathcal{W}}_{n}^{(\nu)}(\tilde{k}) \, \tilde{F}^{\nu} \label{E:PseudoEnergyExp}
\end{equation}
as  power series in $\tilde{F}$. It is assumed that the zeroth order corresponds to the usual Bloch states and band energies. Thus
\begin{equation}
  \tilde{\mathcal{U}}_{n' n}^{(0)}(\tilde{k}) = \delta_{n'n}
\end{equation}
and
\begin{equation}
  \tilde{\mathcal{W}}_{n}^{(0)}(\tilde{k}) = \tilde{\mathcal{E}}_{n}(\tilde{k}).
\end{equation}\\

Replacing the expansions \eqref{E:TransfExp} and \eqref{E:PseudoEnergyExp} in \eqref{E:PseudoEigenvalueTransf} and collecting terms with equal powers of $\tilde{F}$ we can find a recurrent system of equations. To first order it is found that, in the original coordinates, the unitary transformation $U_{n'n}(k)$ is given by \eqref{E:FirstOrderUnitaryTransf} \cite{Adams56, Wannier60}. Note that if $\tilde{F}$ is small we expect the parameter $|\Delta_{n'n}(k)|$, defined in \eqref{E:DeltaParameter} , to be small because the Lax connection is of the order of the lattice constant $b$, while the energy difference is of the order of $E_R$. The first order approximation for the energy $W_n(k)$ is simply the band energy renormalized by the diagonal part of the Lax connection (see equations~\eqref{E:RenormBandEner} and~\eqref{E:FirstOrderEnergy}).

\end{document}